\documentclass[twocolumn,aps,prl,superscriptaddress]{revtex4-1}

\usepackage{color,amsmath,amssymb,epsfig,graphicx}
\usepackage{bm}
\usepackage{mathptmx, courier, pifont}
\usepackage[scaled=0.92]{helvet}
\usepackage[T1]{fontenc}
\usepackage{textcomp}
\usepackage[colorlinks=true,linkcolor=blue,filecolor=blue,urlcolor=blue,citecolor=blue]{hyperref}
\providecommand{\boldsymbol}[1]{\mbox{\boldmath $#1$}}

\usepackage{multirow}
\usepackage{longtable}
\def\bra{\,<\!} \def\ket{\!>\,} \def\ack{\,|\,}
\begin{document}
\author{Nazira Nazir}
\affiliation{Department of Physics, University of Kashmir,
  Hazratbal, Srinagar, 190 006, India}
\author{S. Jehangir}
\email{sheikhahmad.phy@gmail.com}
\affiliation{Department of Physics, Islamic University of Science and Technology,
  Jammu and Kashmir, 192 122, India}
\author{S.P. Rouoof}
\affiliation{Department of Physics, Islamic University of Science and Technology, 
Jammu and Kashmir, 192 122, India}
\author{G.H. Bhat}
\email{gwhr.bhat@gmail.com}
\affiliation{Department of Physics, S.P. College,  Srinagar, Jammu and Kashmir, 190 001, India}
\affiliation{Cluster University Srinagar, Jammu and Kashmir,
  Srinagar, Goji Bagh, 190 008, India}
\author{J. A. Sheikh}
\email{sjaphysics@gmail.com}
\affiliation{Department of Physics, University of Kashmir,
  Hazratbal, Srinagar, 190 006, India}
\author{N. Rather}
\affiliation{Department of Physics, Islamic University of Science and Technology, 
Jammu and Kashmir, 192 122, India}
 \author{ \\S.~Frauendorf}
\email{sfrauend@nd.edu}
\affiliation{Department of Physics and Astronomy, University of Notre Dame, Notre Dame, Indiana 46556,  USA}

\title{ Microscopic aspects of $\gamma$-softness in atomic nuclei }

\begin{abstract}

The microscopic origin of the $\gamma$-softness (fluctuations in the triaxiality parameter $\gamma$ of the nuclear shape)
 observed in atomic nuclei is studied in the framework 
 of the triaxial projected shell model approach, which is based on a multi-quasiparticle
 configuration space generated from a deformed mean-field. It is demonstrated that the coupling to quasiparticle excitations
 drives the system from a $\gamma$-rigid to a $\gamma$-soft pattern. As an  illustrative example for a $\gamma$-soft nucleus,
 a detailed study has been performed for the $^{104}$Ru nucleus. The experimental energies and a large sample of
measured $E2$ matrix elements available for this nucleus are reproduced quite accurately. 
The shape invariant analysis of the calculated $E2$ matrix elements elucidates the  $\gamma$-soft nature of
$^{104}$Ru.
\end{abstract}

\date{\today}

\maketitle

The emergence of  collective phenomena from microscopic degrees of freedom is one of the
 challenging problems in quantum many-body physics. The collective features observed in condensed matter,
atomic, molecular, nuclear and other systems are generally described using macroscopic approaches. In the
case of atomic nuclei,
the concept of a leptodermic droplet is employed and is the basis of the 
collective model introduced by Bohr and Mottelson \cite{BMII}. This model parametrizes the spatial
deformation  of the nuclear density distribution
 in terms of a  multipole expansion. For the interpretation of the low-energy properties, it is often sufficient to consider
only the quadrupole terms  in the multipole expansion, which are  expressed in terms of the
$\beta$- and $\gamma$-shape variables.
These parameters are considered as "collective" dynamical variables,
the motion of which are described  by the Bohr Hamiltonian \cite{BMII,Frauendorf15}.
The terminology of "rigid" and "soft" shape is based on its quantum eigenstates : rigid shape 
signifies small fluctuations around the equilibrium value
and soft shape implies large fluctuations.
As an alternative to the geometric Bohr Hamiltonian, the algebraic 
interacting boson method (IBM) \cite{IBM,Otsuka87} accounts for  the collective quadrupole degrees of freedom
in the framework of a closed Lie algebra of boson operators with "rigid" and "soft" shapes
defined in terms of group theoretical limits \cite{Nomura08, Ben11}. 

The nucleons are in the ballistic regime, i.e., they travel quasi-free inside the nuclear surface and 
due to confinement, the motion is quantized. This generates the shell structure of  the nucleonic states, like 
the shell structure of atoms, the electronic states of molecules or the band structure of the electrons in crystals. 
The quantization of the nucleonic motion governs the 
statics and dynamics of the nuclear shape, i.e., the nucleus may be viewed as a droplet of a fermionic
liquid with long-range correlations, analogous to small metal clusters, He$_3$ droplets,
Rydberg states in large atoms  and ultracold fermionic atom gases.

A self-consistent description of the intertwined dynamics of the shape and single particle degrees of freedom
is one  of the major research themes in nuclear structure physics \cite{Frauendorf15}.
The adiabatic time-dependent mean-field (ATDMF) approach and the generator coordinate method (GCM) \cite{RS80} have become 
the standard approaches to describe the collective dynamics \cite{Frauendorf15,NVR11}. 
The advances in computer technology and the development of the
new algorithms have made the spherical shell model (SSM) a viable approach 
to calculate the collective characteristics of the low-lying states \cite{Alhassid96,Togashi16,Poves20}.

In this work, we propose the triaxial projected shell model (TPSM) \cite{SH99}
as an alternative approach to elucidate the triaxial characteristics of atomic nuclei. In contrast to the
SSM approach, the TPSM uses angular momentum
projected quasiparticle configurations of a triaxial mean field that incorporates essential correlations, and the  residual
correlations  are included through diagonalization of the  Hamiltonian. This drastically reduces
the numerical effort and simplifies the interpretation of the results. We think that such a novel approach may pave a way
 to describe other collective features in atomic nuclei, as well as in non-nuclear
mesoscopic systems consisting of a few hundred particles.


To demonstrate how $\gamma$-softness arises from the TPSM picture, we have chosen
$^{104}$Ru nucleus as an example because a detailed Coulomb excitation (COULEX) experiment has been performed
for  this nucleus with the
measurement of twenty-eight $E2$ and three $M1$ matrix elements \cite{JS06} among the low-lying states.
From the analysis of the shape invariants \cite{DC86,KM72},
derived from the set of $E2$ matrix elements,
it was deduced \cite{JS06} that $^{104}$Ru is "$\gamma$-soft", i.e., the $\gamma$-degree of freedom  is
distributed over a wide range between prolate ($\gamma=0^o$) and oblate ($\gamma=60^o$) shapes.
The authors of Ref. \cite{JS06} compared the observed data with a variant of the  
ATDMF approach \cite{KZ99},
 referred to as  the  quadrupole collective Bohr Hamiltonian (QCBH). 
These calculations reproduced the matrix elements from the COULEX experiment with a good accuracy. 
In particular, the dispersions of the shape invariants indicated that $^{104}$Ru
nucleus is $\gamma$-soft, which  is consistent with the shallow potential of their QCBH. 


In the present work, we  demonstrate  that
experimental energies and the COULEX
matrix elements are accurately reproduced in the framework of the TPSM
approach \cite{SH99}. The results turn out to be quite similar to the ones obtained using the QCBH approach, tabulated
in Ref. \cite{JS06} as the "dynamic" (dyn.) variant.
In particular, the important features of $\gamma$-softness are obtained without considering a collective Hamiltonian,
instead they emerge directly by mixing multi-quasiparticle configurations with a fixed deformation.

It was demonstrated in Ref. \cite{JS21}, for a large set of nuclei, that TPSM approach reproduces the experimental staggering
parameter 
\begin{equation}
 S(I) = \frac{[E(I)-E(I-1)]-[E(I-1)-E(I-2)]}{E(2^{+}_1)}  ,
\end{equation}
of the $\gamma$-bands, which allowed  to 
classify the studied nuclei as $\gamma$-soft or $\gamma$-rigid. 
In this work, we demonstrate for the first time that the TPSM reproduces the more direct criteria for
$\gamma$-softness by means of the shape invariants deduced from the $E2$ transition matrix elements.
This method has also been used to extract the collective quadrupole characteristics from 
large-scale SSM calculations  \cite{Poves20}. 


The basic methodology of the TPSM approach is similar to the SSM
with the exception that angular-momentum projected deformed basis
is employed to diagonalize the shell model Hamiltonian \cite{KY95}.
The Hamiltonian consists of monopole pairing ($ \hat P^\dagger \hat P$), quadrupole pairing ($\hat
P^\dagger_\mu\hat P^{}_\mu$), and
quadrupole-quadrupole ($\hat Q^\dagger_\mu
\hat Q^{}_\mu$) interaction terms within the configuration space of three major oscillator shells
($N=3,4,5$ for
neutrons and $N=2,3,4$ for protons) $:$ 
\begin{equation}
\hat H = \hat H_0 - {1 \over 2} \chi \sum_\mu \hat Q^\dagger_\mu
\hat Q^{}_\mu - G_M \hat P^\dagger \hat P - G_Q \sum_\mu \hat
P^\dagger_\mu\hat P^{}_\mu ,
\label{hamham}
\end{equation}
where $\hat H_0$  is the spherical single-particle potential \cite{Ni69}.
The pairing parameters are taken from our earlier work \cite{JS16} 
with $G_M$ chosen such that the overall observed odd-even mass differences are
reproduced for nuclei in this region, and the quadrupole pairing strength
$G_Q$ is assumed to be 0.18 times $G_M$. The QQ-force
strength $\chi$ is fixed by the self-consistency 
relation between the input deformation and the quadrupole mean field  (see Ref. \cite{KY95}).

The shell model Hamiltonian of Eq. (\ref{hamham}) is diagonalized in the space of angular-momentum projected
multi-quasiparticle states. 
In the present work, the 
basis is composed of the 0-qp vacuum, the two-quasiproton, the two-quasineutron and the combined
four-quasiparticle configurations
\begin{eqnarray}\label{basis}
&&\hat P^I_{MK}\ack\Phi\ket,~
\hat P^I_{MK}~a^\dagger_{p_1} a^\dagger_{p_2} \ack\Phi\ket,~
\hat P^I_{MK}~a^\dagger_{n_1} a^\dagger_{n_2} \ack\Phi\ket,\nonumber\\
&&\hat P^I_{MK}~a^\dagger_{p_1} a^\dagger_{p_2}
a^\dagger_{n_1} a^\dagger_{n_2} \ack\Phi\ket,
\end{eqnarray}
respectively, where $a^\dagger$ is the quasiparticle creation operator, $''p''$ labels the quasiproton and $''n''$ the quasineutron states,
$P^I_{MK}$ is the standard three-dimensional angular-momentum projection operator \cite{RS80},
and $\ack\Phi\ket$ represents the triaxially-deformed quasiparticle  vacuum state.

The electromagnetic transition matrix elements are calculated  using the electric quadrupole tensor ${\cal M}( E2)_\mu$ 
with the effective charge of $1.5e$ for protons and  $0.5e$ for neutrons, and the magnetic dipole operator  ${\cal M}( M1)_\mu$
with the spin $g$ factors reduced by a factor of $0.75$ \cite{sj17}.
The deformed
basis states are generated from a mean-field Nilsson Hamiltonian with fixed deformation parameters 
 $\epsilon=0.258$ and $\epsilon'=0.150$, which corresponds to a triaxility parameter
of $\gamma=\arctan(\epsilon'/\epsilon)=30.2^o$.
As in our earlier studies \cite{sj17,JS21b,JS21}, 
 these two parameters were adjusted to reproduce the $B(E2, 2^+_1\rightarrow 0^+_1)$ value \cite{raman}
 and the energy  $E(2^+_2)$ of the $\gamma$-band of $^{104}$Ru.  Moreover, the pairing
 strength was reduced by a factor of 0.5 for the bands built on 
the excited $0^+_2$ and $0^+_3$ states.

\begin{figure}[htb]
 \centerline{\includegraphics[trim=0cm 0cm 0cm
0cm,width=0.55\textwidth,clip]{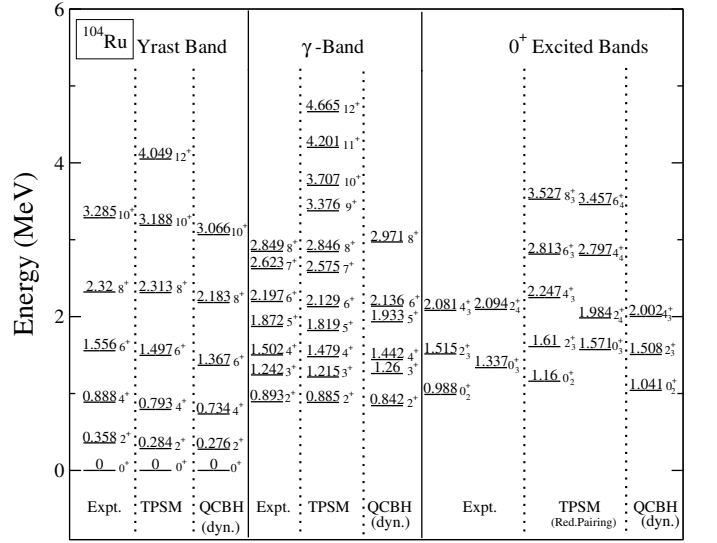}} 
\caption{ Comparison of the measured energy levels of  $^{104}$Ru nucleus  with the  results of TPSM  and QCBH calculations.
The QCBH energies are taken from Fig. 6 of Ref. \cite{JS06} level sequence QCBH(dyn.).
}
\label{fig:energies}
\end{figure}

\begin{figure}[htb]
 \centerline{\includegraphics[trim=0cm 0cm 0cm 0cm,width=0.52\textwidth,clip]{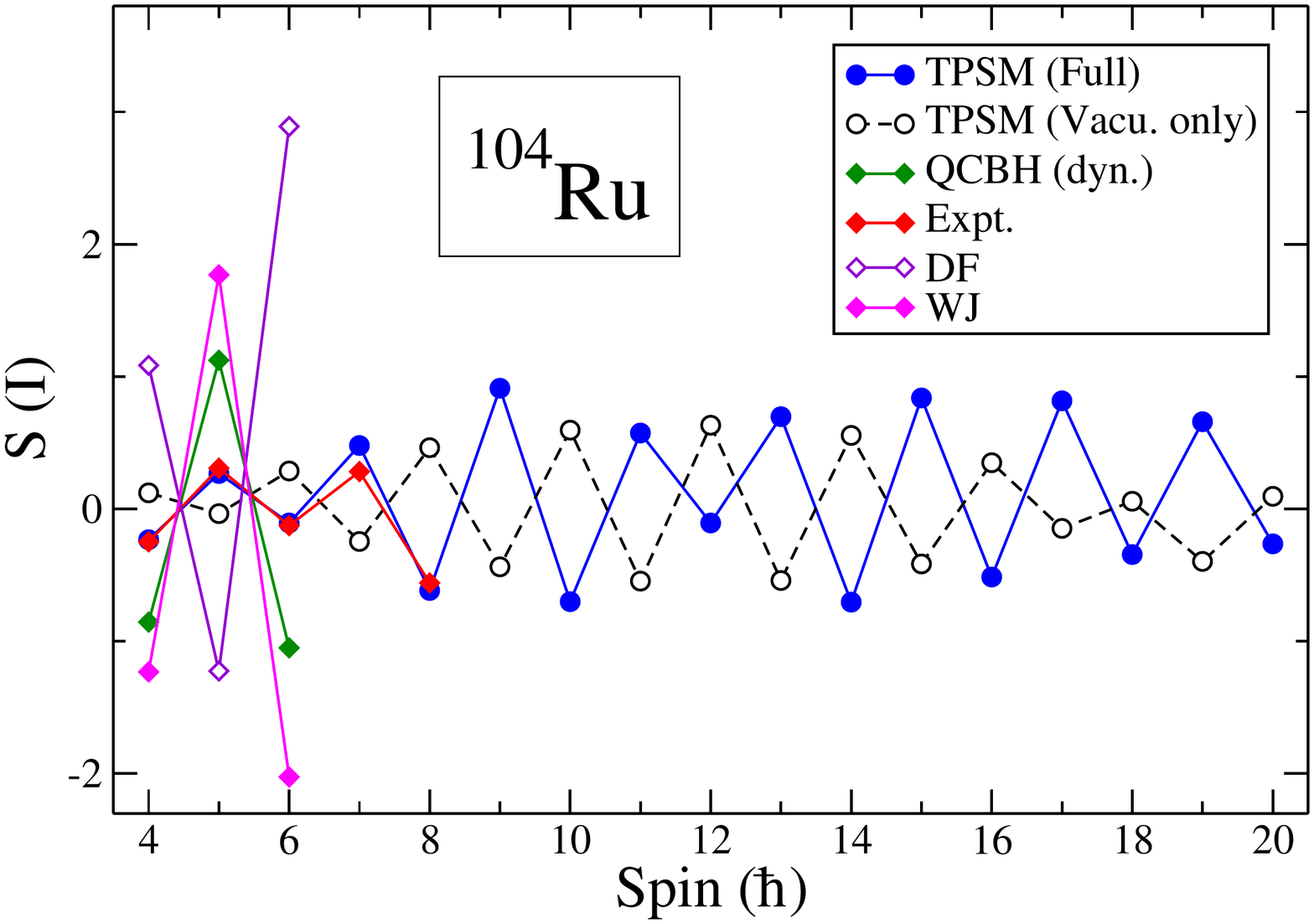}} 
\caption{(Color
online) Staggering parameter $S(I)$  for the $\gamma$-band in $^{104}$Ru. The criterion for $\gamma$-softness is  $S(I_{even})<S(I_{odd})$ 
and $\gamma$-rigidness is $S(I_{odd})<S(I_{even})$.
The observed  values (Expt.) are compared with the TPSM values including the quasiparticle excitations (Full) , 
the TPSM with vacuum configuration only (Vacu. only), the QCBH (dyn.) values, the Davydov Filippov (DF) and the Wilets Jean (WJ) limits. 
The QCBH (dyn.), DF and  WJ staggering values have been calculated from the energies of the
respective models, shown in Fig. 6 of Ref. \cite{JS06}.
} 
\label{fig:staggering}
\end{figure}

Fig. \ref{fig:energies} compares the TPSM energies  with the experimental data and also with the corresponding energies obtained
 in the QCBH approach. It is evident from the figure that the TPSM reproduces the experimental energies quite well. 
Fig. \ref{fig:staggering} displays the staggering parameter
 $S(I)$ obtained from the corresponding energies of the $\gamma$-band. 
 The  relation between staggering pattern of $\gamma$ softness  were reviewed in Ref. \cite{Frauendorf15}.
The experimental $S(I)$ depicts the even-I-down pattern of a $\gamma$-soft nucleus, which is very well reproduced by the TPSM 
calculations that include the quasiparticle admixtures. The TPSM results with the vacuum configuration 
only show the odd-I-down pattern of a $\gamma$-rigid nucleus. The feature of the $\gamma$-softness is, 
therefore, generated by the quasiparticle admixtures.
The QCBH values indicate $\gamma$-softness as well, however, overestimate the amplitude of $S(I)$.

\begin{table}[h!]
\LTcapwidth=0.4\textwidth
\caption{Comparison of all known experimental reduced  $E2$ diagonal, in-band and inter-band matrix elements 
 $\bra I_i || E2 || I_f\ket(e.b)$, (associated errors in parenthesis) and calculated ones for yrast- and $\gamma$-band of
    $^{104}$Ru.}
\resizebox{0.45\textwidth}{!}
  {
\begin{tabular}{|c|c|c|c||c|c|c|c|}
  \hline
  $I_i \rightarrow I_f$   & Expt.    &TPSM       &TPSM
  & $I_i \rightarrow I_f$ & Expt.    &TPSM       &TPSM
  \\
                          &         &(Full)     &(Vacu.)
  &                       &         &(Full)     &(Vacu.)                
  \\

  \hline
  $2_1\rightarrow 2_1$    &-0.71(\textit{11)}             &-0.817  &-0.634
  &$4_2\rightarrow 3_1$   &$\pm $0.68  (\textit{5)}       &-0.787  &-0.597 \\
  $4_1\rightarrow 4_1$    &-0.79(\textit{15)}             &-0.906  &-0.437
  &$5_1\rightarrow 3_1$   &1.22(\textit{4)}               &1.184   &0.697\\
  $6_1\rightarrow 6_1$    &-0.70(\textit{$^{+30}_{-20}$ })  &-0.868  &-0.342
  &$6_2\rightarrow 4_2$   &1.52 (\textit{12)}             &1.521   &0.682\\ 
  $8_1\rightarrow 8_1$    &-0.6(\textit{$^{+3}_{-5}$ })     &-0.855  &-0.297
  &$8_2\rightarrow 6_2$   &2.02(\textit{4)}               &2.056   &0.747\\
  $2_2\rightarrow 2_2$    &0.62(\textit{8)}               &0.648   &0.633
  &$2_2\rightarrow 0_1$   &-0.156 (\textit{2) }           &-0.141  &-0.225\\
  $4_2\rightarrow 4_2$    &-0.58(\textit{18)}             &-0.749  &-0.534
  &$2_2\rightarrow 2_1$   &-0.75(\textit{4) }             &-0.722  &-0.612\\
  $6_2\rightarrow 6_2$    &$\pm$1.0(\textit{3)}           &-1.105  &-0.763
  &$2_2\rightarrow 4_1$   &$\epsilon$ [-0.1, 0.1]         &-0.090  &-0.001\\
  $2_1\rightarrow 0_1$    &0.917 (\textit{25)}            &0.973   &0.901
  &$3_1\rightarrow 2_1$   &0.22(\textit{10)}              &0.254   &0.302\\ 
  $4_1\rightarrow 2_1$    &1.43 (\textit{4)}              &1.591   &1.456
  &$3_1\rightarrow 4_1$   &-0.57                          &-0.517  &-0.559\\
  $6_1\rightarrow 4_1$    &2.04 (\textit{8)}              &2.081   &1.830  
  &$4_2\rightarrow 2_1$   &-0.107 (\textit{8 ) }          &-0.113  &-0.054 \\  
  $8_1\rightarrow 6_1$    &2.59 (\textit{$^{+24}_{-9}$ )}   &2.486   &1.902
  &$4_2\rightarrow 4_1$   &-0.83(\textit{5) }             &-0.840  &-0.505 \\
  $10_1\rightarrow 8_1$   &2.7 (\textit{6)}               &2.668   &1.623
  &$6_2\rightarrow 4_1$   &-0.22(\textit{$^{+6}_{-12}$ })   &-0.230  &-0.682\\
  $3_1\rightarrow 2_2$    &-1.22 (\textit{10)}            &-1.241  &-0.935
  &$6_2\rightarrow 6_1$   &$>$-0.84                       &-0.947  &-0.411 \\
  $4_2\rightarrow 2_2$    &1.12(\textit{5)}               &1.095   &0.510
  &                      &                               &        &      \\

  \hline

\end{tabular}
}
\end{table}

\begin{table}[h!]
\LTcapwidth=0.4\textwidth
\caption{Comparison of all known experimental reduced  $E2$ matrix elements 
 $\bra I_i || E2 || I_f\ket(e.b)$, diagonal, in-band  and inter-band  values (associated errors in parenthesis) and calculated ones for excited $0^+$ bands of
    $^{104}$Ru.}
\resizebox{0.4\textwidth}{!}
  {
\begin{tabular}{|c|c|c||c|c|c|}
  \hline
$I_i \rightarrow I_f$ & Expt.    & TPSM       &$I_i \rightarrow I_f$ & Expt.  & TPSM   \\
  \hline
  $2_3\rightarrow 0_2$    &0.71(\textit{4)}                 &0.682   
  &$2_3\rightarrow 4_1$   & -0.370(\textit{4)}              &-0.311\\
  $4_3\rightarrow 2_3$    &0.75(\textit{25)}                &0.613  
  &$2_3\rightarrow 2_2$   &$\pm$0.22(\textit{$^{+25}_{-5}$})  &-0.237\\  
  $0_2\rightarrow 2_1$    &-0.266(\textit{8)}               & -0.221
  &$2_3\rightarrow 4_2$   &0.31(\textit{$^{+13}_{-6}$})       & 0.221\\ 
  $0_2\rightarrow 2_2$    &0.08 (\textit{3)}                & 0.099
  &$2_3\rightarrow 4_4$   &0.53(\textit{$^{+32}_{-14}$})      &0.481  \\ 
  $2_3\rightarrow 0_1$    &-0.071(\textit{3)}               &-0.048  
  &$0_3\rightarrow 2_1$   & $>$-0.1                         &-0.201 \\ 
  $2_3\rightarrow 2_1$    & $\pm$0.07(\textit{3)}           &-0.031
  &$2_3\rightarrow 2_3$    &-0.08(\textit{$^{11}_{25}$)}      &-0.631\\  

\hline
\end{tabular}
}
\end{table}

\begin{table}[h!]
\LTcapwidth=0.4\textwidth
\caption{Comparison of all known experimental reduced  $M1$ matrix elements 
 $\bra I_i || M1 || I_f\ket(\mu_N)$, in-band and
inter-band values (associated errors in parenthesis) and calculated ones for
$^{104}$Ru.}
\resizebox{0.4\textwidth}{!}
  {
\begin{tabular}{|c|c|c||c|c|c|}
  \hline
  $I_i \rightarrow I_f$      & Expt.                        & TPSM
  &$I_i \rightarrow I_f$     & Expt.                        & TPSM\\
\hline
$2_1\rightarrow 3_1$        & -0.054(\textit{$^{-9}_{+9}$})   & -0.044
&$2_1\rightarrow 2_1$       & 0.82(\textit{10})             & 0.791\\
$2_1\rightarrow 2_2$        &$<0.02$                        & -0.038
&$4_1\rightarrow 4_2$       &-0.15(\textit{$^{-3}_{+3}$})     & -0.136\\

   \hline
\end{tabular}
}
\end{table}
\begin{figure}[htb]
 \centerline{\includegraphics[trim=0cm 0cm 0cm
0cm,width=0.45\textwidth,clip]{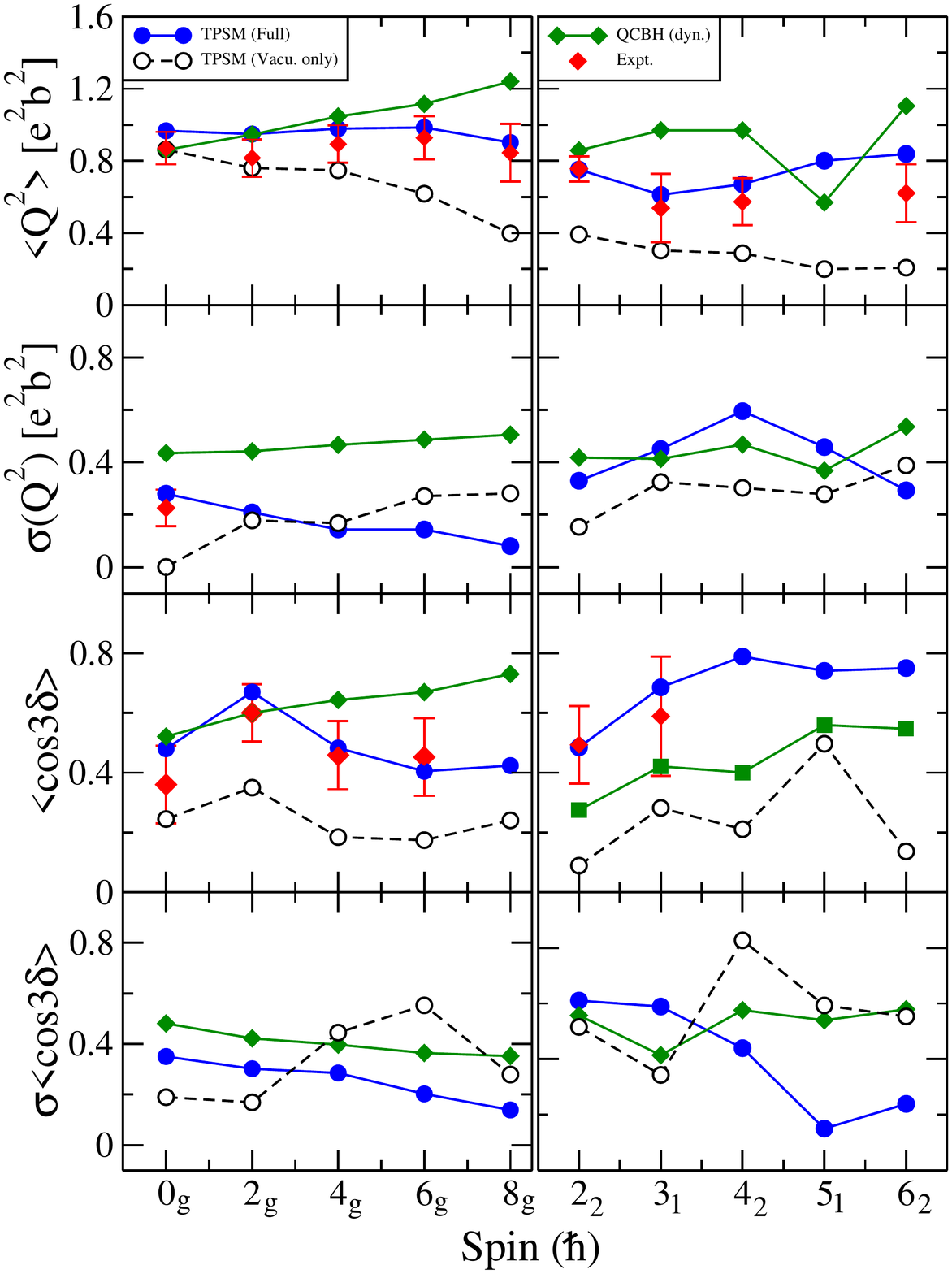}}
 \caption{(Color
online) Comparison of the observed (Expt.), TPSM , and QCBH (dyn.)  shape invariants
for the ground-band (left panels) and $\gamma$-band (right panels) in $^{104}$Ru. 
The TPSM quantities in this figure and in Fig. \ref{fig:invariants0p} 
were calculated using the code GOSIA \cite{GOSIA2012}, 
and the QCBH (dyn.) values were taken from Figs. 4 and 5 of Ref. \cite{JS06}. 
} 
\label{fig:invariants}
\end{figure}


Tables I and II compare the experimental reduced $E2$ matrix elements with the ones calculated using the TPSM wavefunctions.
 The corresponding values of the QCBH calculations are given in the Tables 1-3 of Ref. \cite{JS06}. 
We adopted the phase convention of Ref. \cite{JS06} (Section 6).
 The agreement of the TPSM results with the COULEX data and QCBH calculations is remarkable, 
 because the $E2$ matrix elements provide the most direct information on the statics and dynamics of the collective quadrupole
 modes.  Comparing with the matrix elements Vacu. shows that the quasiparticle admixtures generate the substantial shifts needed for
 the good agreement with the experiment.  
 Table III presents  TPSM  reduced  $M1$ matrix elements,
 which also reproduce the known experimental values quite well.
 
Fig. \ref{fig:invariants} depicts the quadrupole shape invariants and their dispersions calculated from the reduced $E2$ matrix elements.
 The pertaining expressions are given in  Refs. \cite{JS06,DC86,KM72} . 
 The invariant $\bra I^+_n | Q^2 | I^+_n \ket$ measures the average intrinsic deformation of a state $I^+_n$. 
 The invariant $\bra I^+_n |\cos 3 \delta | I^+_n \ket$ contains the information about the triaxiality of the intrinsic shape, 
 where $\delta=\arctan{\bra I^+_n |\cos 3 \delta | I^+_n \ket}$.
 The parameters $Q$ and $\delta$ can be related to the effective deformation parameters
  $\beta$ and $\gamma$ of the liquid drop model \cite{BMII}. The relevant formulae for ellipsoidal 
  shape are given in the appendix of Ref. \cite{JS06}. 
  Evaluating them, we found $\delta<\gamma$ with $\vert \delta-\gamma\vert <2.5^\circ$.

 The TPSM values $\bra Q^2\ket \approx 1.0 (eb)^2$ for the ground-band are nearly constant in agreement with the experimental data.
 The corresponding effective deformation of $\beta=0.28$ is the input in the TPSM.
  The TPSM values of $\bra Q^2\ket\sim0.7 $ corresponding to  $\beta=0.27$
  indicate a smaller deformation for the $\gamma$-band,
 which  is again consistent with the experiment. 
  
 For the ground-band, the TPSM values of $\bra \cos 3 \delta\ket$ signify a substantial triaxiality with preference for
 prolate shape. The value of 0.6 corresponds to $\delta = 18^\circ$. For $I=2$, 
 the experimental data shows a shift towards prolate shape, which is seen in the TPSM values as well. 
   The TPSM dispersion $\sigma<\cos 3 \delta> \sim 0.3$  signifies large 
  fluctuations of the triaxiality parameter with 70\% of the distribution within the range $9^\circ < \delta<24^\circ$. 
  This
  indicates that $^{104}$Ru is $\gamma$-soft in accordance with the staggering parameter $S(I)$ of Fig. \ref{fig:staggering}.
  The QCBH dispersion of $\sigma<\cos 3 \delta> \sim 0.4$ corresponds to a distribution with 70\%  in the larger  range  
  $0^\circ < \delta<26^\circ$. 
  The QCBH values seem to overestimate the $\gamma$-softness, which is consistent with the QCBH values for $S(I)$ being larger than the TPSM  and the experimental values.
  
  For the $\gamma$-band, the TPSM values of $\bra \cos 3 \delta\ket$ for $I=2,~3 $ and 4 indicate a shift towards prolate shape,
  which is in accordance with the experimental values. This shift is also seen in the QCBH calculation, 
  which predicts a larger $\delta\sim  24^\circ$ as compared to $\delta\sim  18^\circ$ of TPSM. The dispersions are similar to the ones in the
  ground-band.
  
  
\begin{figure}[htb]
\vspace{1cm}
\includegraphics[width=0.5\textwidth]{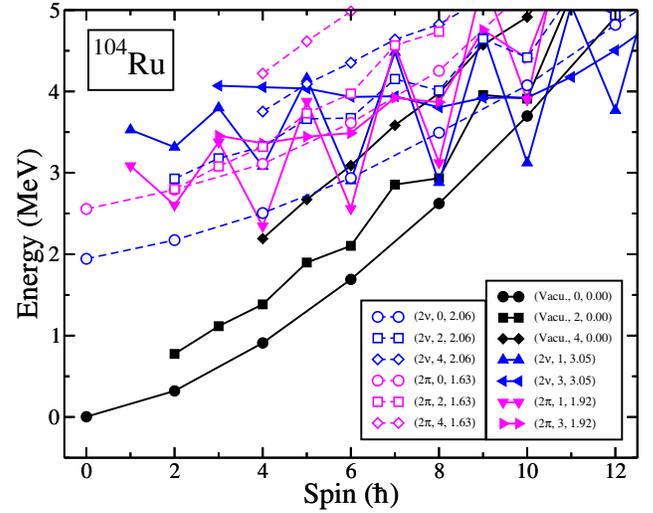}
\caption{(Color online) TPSM projected energies before band mixing. The bands are labelled by
three quantities : quasiparticle character, $K$-quantum number and energy of the two-quasiparticle
state. For instance, $(2\nu,1,3.05)$ designates the $K=1$ state projected from
the $h_{11/2}$ two-quasineutron configuration with the energy of 3.05 MeV.
 The  $K=0, ~2,~4$ states projected from the quasiparticle vacuum are labelled with Vacu.
 The four-quasiparticle states lie above 5 MeV.} \label{fig:Banddiag}
\end{figure}
\begin{figure}[htb]
 \centerline{\includegraphics[trim=0cm 0cm 0cm 0cm,width=0.5\textwidth,clip]{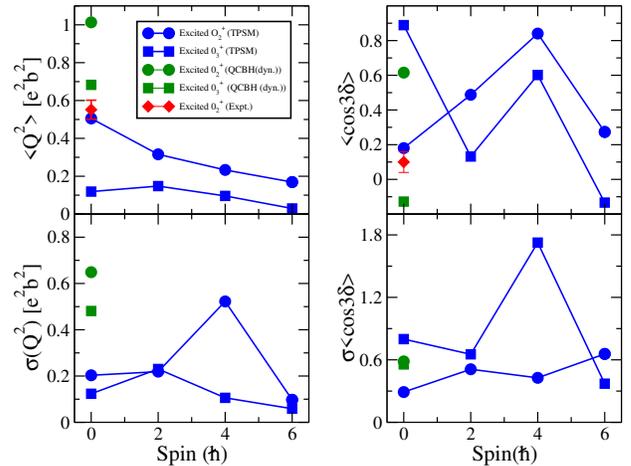}}
 \caption{(Color
online) Comparison of the observed (Expt.), QCBH (dyn.) and TPSM  shape invariants
for the excited $0^+$ bands in $^{104}$Ru. 
} 
\label{fig:invariants0p}
\end{figure}


In order to probe the microscopic origin of the appearance of $\gamma$-soft characteristics,
the energies of the  angular momentum projected  quasiparticle configurations are
depicted in Fig. \ref{fig:Banddiag}. The $K=0,2,4,.....$
states, projected from the quasiparticle vacuum configuration, have an increasing content of the angular momentum oriented along
the long-axis with the smallest moment of inertia. 
For $K=0$, there are only even-$I$ states. When the TPSM Hamiltonian is diagonalized with vacuum configuration only, it
mostly couples the $K=0$ with the $K=2$ states. The coupling pushes up
   the even-$I$ states of the $\gamma$-band above the mean energy of the adjacent $K=2$ states with odd $I$, which are not
   shifted because they do not have a partner in
   the ground-band.  This results in the odd-$I$ low pattern of the staggering parameter $S(I)$.
   
  In the TPSM diagonalization, the most important quasiparticle admixtures to the $\gamma$-band are the $K=1$ state projected from 
  the $h_{11/2}$  two-quasineutron configuration and the $K=1$, $3$ states, projected from the $g_{9/2}$ two-quasiproton  
  configuration. Their admixtures increase with  $I$ and are about 20\% for the $I=7$ state.  As noted from
  Fig. \ref{fig:Banddiag}, these quasiparticle band structures
  show a pronounced even-$I$ low staggering. The coupling pushes the  states  with even $I$ further down than the states
  with odd $I$, and the coupling is strong enough to reverse the staggering pattern of the $\gamma$-band into the even-$I$ low pattern of a 
  $\gamma$-soft nucleus. 
 The high-j orbitals $h_{11/2}$ and $g_{9/2}$ have large quadrupole moments, and a moderate admixture of these states modifies the  $E2$ 
 matrix elements such that the pattern of a $\gamma$-soft nucleus emerges from the $\gamma$-rigid pattern of   
the vacuum configuration only. The superposition of configurations with different triaxiality increases the shape fluctuations as seen in Fig. \ref{fig:invariants}.

A further indication for $\gamma$-softness is the appearance of a $\Delta I=2$ band built on an excited $0^+_2$ state which is connected by
 collective $E2$ transitions with the $\gamma$-band. In terms of the collective Bohr Hamitonian, it represents a pulsation
 of the triaxial surface that carries no angular momentum (see Fig. 6 of the review \cite{Frauendorf15}).  
 The limiting cases are the $K=0$ double $\gamma$ vibration about the axially symmetric equilibrium shape \cite{BMII}  and the
  $0^+_2$ state in the $\gamma$ -independent potential, where it represents the oscillation between prolate and oblate shapes through a triaxial path.
 The COULEX experiment observed $0^+_2$ and $0^+_3$ states of this type \cite{JS06}. As seen in Fig. \ref{fig:energies} and Table II, the
  TPSM  calculations account well for the experimental data.  For the $0^+_2$ state, the shape invariant $\bra \cos 3 \delta\ket$ in Fig. \ref{fig:invariants0p}
  indicates strong triaxiality of $\delta=26^\circ$, 
  which decreases towards the prolate value of zero with $I$. For the $0^+_3$ state, weak triaxiality is noted that increases to the maximal value with $I$.
  The dispersions indicate large fluctuations in $\delta$. The $0^+_2$ state contains a 60\% component of the $K=0$ two $h_{11/2}$ quasineutron 
  configuration (2$\nu$, 0, 2.06 in Fig. \ref{fig:Banddiag}), which decreases to 20\% for the $6^+_3$ state by admixing  many small components.
  The $0^+_3$ state contains a 95\% component of the $K=0$ two $g_{9/2}$ quasiproton 
  configuration (2$\pi$, 0, 1.63 in Fig. \ref{fig:Banddiag}), which decreases to 40\% for the $6^+_4$ state by admixing many small components.
  
  The dominating two-quasiparticle character of the $0^+_{2,3}$ states justifies the reduction of the pairing strength in the TPSM calculation,
  which lie too high in the excitation energy with full strength
  (see Fig. \ref{fig:Banddiag}). The QCBH calculations \cite{JS06} also point to a change of the pair correlations. 
 
To conclude, we have
shown that TPSM approach reproduces the extended set of collective $E2$ matrix elements measured in the COULEX experiment
with a remarkable accuracy. The shape invariant quantities deduced from these matrix elements capture the features of a $\gamma$-soft nucleus with the mean value of 
the triaxiality parameter of $\delta\sim  20^\circ$, and fluctuations within the range $0^\circ < \delta<26^\circ$. The important inference drawn 
from the present analysis is that admixtures of quasiparticle states into the collective vacuum configuration can transform the pattern of the energies and  
 $E2$ matrix elements  from $\gamma$-rigid to that of $\gamma$-soft character. 
The spherical shell model calculations with a sufficiently large basis set should also capture the discussed collective features \cite{Togashi16,Alhassid96,Poves20}.
 However, the small number of
 quasiparticle configurations in the present deformed shell model approach provides another perspective on the nature of
 $\gamma$-softness in atomic nuclei : it represents the superposition of a few configurations with different triaxiality.
 
 \section{ACKNOWLEDGEMENTS}
The authors  acknowledge the Science and Engineering Research Board (SERB), Department of Science and
Technology (Govt. of India) for providing financial assistance under the 
Project No.CRG/2019/004960, and for the award of INSPIRE fellowship to one of the author (NN). We are also indebted to
Drs. L. Pr\'ochniak, P. J. Napiorkowski and J. Srebrny for providing some unpublished results, 
and for their assistance in running the GOSIA code.
\bibliographystyle{apsrev4-1}
\bibliography{odd_Xe_TPSM}
\end{document}